\documentclass[reqno,11pt]{amsart}

\usepackage{amssymb,amsthm}
\usepackage{amsmath}
\usepackage{amsfonts}
\usepackage{dsfont}

\usepackage[a4paper,  margin=3cm]{geometry}

\usepackage[active]{srcltx}
\makeatletter\@addtoreset{equation}{section}\makeatother

\newtheorem{theorem}{Theorem}[section]
\newtheorem{corollary}[theorem]{Corollary}

\newtheorem{proposition}[theorem]{Proposition}
\newtheorem{assumption}[theorem]{Assumption}

\theoremstyle{remark}

\numberwithin{equation}{section}

\title[On the Lee-Yang property of some ferromagnets]{On the Lee-Yang property of some ferromagnets}

\author{ Yuri  Kozitsky}

\address{Instytut Matematyki, Uniwersytet Marii Curie-Sk{\l}odowskiej, 20-031 Lublin, Poland}
\email{jurij.kozicki@mail.umcs.pl}


\begin{document}

\begin{abstract}
According to the Lieb-Sokal theorem, the partition  function, $Z$, of a ferromagnetic spin model has the Lee-Yang property if the single-spin partition function has it. In this note, it is shown that for some spin models a ferromagnetic interaction can induce the Lee-Yang property of $Z$ even if the single-spin partition  function fails to have it. In particular, this holds for the Blume-Capel model and for the annealed states of the $s=\pm 1$ site dilute Ising model with a neares-neighbor interaction on $\mathds{Z}^d$, as well as with interactions defined by a hierarchical structure similar to that of Dyson's hierarchical model.

\end{abstract}

\maketitle

\section{Introduction}
According to the celebrated Lee-Yang theorem, the partition function of the ferromagnetic spin $s=\pm 1$ Ising model can be written in the form 
\begin{equation}
 \label{1}
 Z (h) = Z(0) \prod_{j=1}^\infty (1+ \gamma_j (\beta h)^2),    
\end{equation}
where $h$ is an external magnetic field, $\beta$ is the inverse temperature and the parameters $\gamma_j$, $j\in \mathds{N}$, satisfy
\begin{equation}
 \label{2}
 0 < \gamma_{j+1} \leq \gamma_j,  \qquad \sum_{ j=1}^\infty \gamma_j < \infty.
\end{equation}
In this case, the partition function -- and the model itself -- are said to have the Lee-Yang property.
By \eqref{1} it follows that $Z(h) = 0$ for $\beta h = \pm x_j := \pm i / \sqrt{\gamma_j}$, $j\in \mathds{N}$, and thus $i / \sqrt{\gamma_1}$ is  the `first Lee-Yang zero' which moves towards the origin at the critical point \cite{JN1}. In view of this, the Lee-Yang theorem contributes to the collection of powerful methods of studying phase transitions in the Ising and similar spin models, see \cite{Bena,JN1} and the literature quoted therein. 

The single-spin partition function of the $s=\pm 1$ Ising model, 
\begin{equation}
 \label{2a}
 Z_1 (h) = \sum_{\sigma = \pm 1} e^{\beta h \sigma} = 2 \phi (\beta h), \quad \phi(x) := \cosh (x),
\end{equation}
can also be written in the form of \eqref{1}. A natural generalization of the Lee-Yang theorem might be claiming the validity of \eqref{1}, \eqref{2} for $Z$ also in the case of single-spin distributions other than that in \eqref{2a} -- including those for `unbounded' spins \cite{LP}, being probability measures on the real line. 
Such models are used in the Euclidean quantum field theory, see \cite{GJ}, and in the statistical mechanics of anharmonic crystals, see \cite{BLL,LP}. 
This generalization was performed by C. M. Newman in \cite{Newman}, and then by E. Lieb and A. Sokal in \cite{LiebS}. The method of the latter work is based on the following representation 
\begin{equation}
 \label{3}
Z_N(h) = 2^N \left[ \exp\left( \frac{1}{2} \sum_{i,j=1}^N \beta J_{ij} D_i D_j) \right) \prod_{i=1}^N \phi(x_i) \right]_{{\rm all} \ x_i = \beta h}, \quad D_i = \frac{\partial}{\partial x_i},
\end{equation}
where $\phi$ is the analog of that in \eqref{2a}, related to the aforementioned probability measure $\chi$ by
\begin{equation}
 \label{4}
 \phi(x) = \int_{\mathds{R}} e^{x\sigma} \chi (d\sigma).
\end{equation}
The infinite order differential operator in \eqref{3} can rigorously be defined for all $J_{ij} \geq 0$ if  $\phi$ is an entire function of order less than two, or of order two and of minimal type. In particular, one can take $\phi$ in the form 
\begin{equation}
 \label{5}
 \phi(x) = \psi(x^2) = \prod_{j=1}^\infty (1+\gamma_j x^2), 
\end{equation}
where the collection $\{\gamma_j\}$ satisfies \eqref{2}, and hence $\psi$ is a Laguerre entire function, see \cite{Koz}. 

In the sequel, we crucially use relevant results of \cite{LiebS}, which we formulate now in the form adapted to the present context.
\begin{proposition}\cite[Proposition 2.2, page 157]{LiebS}
 \label{0pn}
 Let $G$ be an exponential type entire function of $(x_1, \dots , x_N)\in \mathds{C}^N$ and $F$ be defined by the formula
 \begin{equation}
  \label{5a}
 F(x_1, \dots , x_N) = \exp\left(\frac{1}{2} \sum_{i,j=1}^N K_{ij} D_i D_j \right) G(x_1 , \dots, x_N), \quad  D_i = \frac{\partial}{\partial x_i}.
\end{equation}
If $G(x_1 , \dots, x_N) \neq 0$ whenever $\Re x_i >0$ and $K_{ij}\geq 0$ for all $i,j=1, \dots N$, then  $F(x_1 , \dots, x_N) \neq 0$ whenever all $\Re x_i >0$.
 \end{proposition}
Its direct corollary is the following statement.
\begin{proposition} \cite[Corollary 3.3, page 165]{LiebS}
 \label{1pn} 
Let $\phi$ in \eqref{3}  be as in \eqref{5} and $J_{ij}\geq 0$ for all $i,j =1, \dots , N$. Then $Z_N(h)$  has the Lee-Yang property.
\end{proposition}
In other words, $Z_N$ has the Lee-Yang property for all $N$ if it has it for $N=1$. Examples of single-spin measures $\chi$ possessing the Lee-Yang property are given in \cite{Koz}.

By the very formulation of Proposition \ref{1pn} it follows that the condition imposed on $\phi$ (hence on $\chi$) is a sufficient one. Thus, one may expect that for certain spin models a ferromagnetic interaction might induce the Lee-Yang property of $Z_N$, $N\geq 2$, even if the single-spin distribution fails to have it.  In this note,  examples of such models are given. 
To the best of our knowledge, this is the first result of this kind.

\section{The Models and the Result}

Here we describe two spin models which we are going to deal with.

\subsection{The models}

 The Blume-Capel model, see, e.g., \cite{Saul}, is the spin model with $s=\pm 1, 0$, the Hamiltonian of which contains the single-ion unisotropy term and is taken in the form
\begin{equation}
 \label{6}
 H_N = - \frac{1}{2}\sum_{i,j=1}^N J_{ij} \sigma_i \sigma_j + \Delta \sum_{i=1}^N \sigma_i^2 - \sum_{i=1}^N h \sigma_i.
\end{equation}
This model has a number of interesting properties, which are essentially different from those of the classical $s=\pm1$ Ising model. In particular, the model demonstrates a tricritical behavior, see \cite{Malakis,leila,Saul,Silva}. In view of the additivity of the unisotropy term in \eqref{6}, one can include $e^{-\beta \Delta \sigma_i}$ in the single-spin distribution $\chi$. In this case, by \eqref{4} it follows that, cf. \eqref{2a},  
\begin{gather}
 \label{7}
Z_1 (h) = 2 e^{-\beta \Delta} \cosh(\beta h) + 1 =: (1+ 2 e^{-\beta \Delta}) \phi(\beta h), \\[.2cm] \nonumber \phi(x) = \frac{\cosh(x) + \theta}{1 +\theta}, \qquad \theta = e^{\beta\Delta}/2.
\end{gather}
Clearly, the latter $\phi$ has the representation as in \eqref{5} only for $\theta \leq 1$. Therefore, Proposition \ref{1pn} guarantees that the ferromagnetic Blume-Capel model has the Lee-Yang property whenever 
\begin{equation}
 \label{8}
 \Delta \leq \beta^{-1} \ln 2.
\end{equation}
The Blume-Capel model has the following natural analog. Assume that the single-spin distribution is random; for instance, $h$ has a random additive part, or the corresponding magnetic particles are thinned out at random. In the physical literature, the latter case is reffered to as a dilute ferromagnet, see, e.g., \cite{Hol,Rush}. In the annealed states, the partition function of the spin  $s=\pm 1$ Ising model has the form, cf. \eqref{3},
\begin{equation}
 \label{9}
Z_N(h) = 2^N \left[ \exp\left( \frac{1}{2} \sum_{i,j=1}^N \beta J_{ij} D_i D_j) \right)\bigg{\langle} \prod_{i=1}^N \phi(x_i) \bigg{\rangle} \right]_{{\rm all} \ x_i = \beta h},
\end{equation}
where $\langle \cdot \rangle$ denotes averaging with respect to the mentioned randomness. For the Bernoulli thinning, the magnetic particles are deleted independently with probability $q= 1-p$. Then
\begin{equation*}
 \bigg{\langle} \prod_{i=1}^N \phi(x_i) \bigg{\rangle} = \prod_{i=1}^N \langle \phi(x_i) \rangle =: \prod_{i=1}^N \varphi(x_i), \quad \varphi(x) =  p \cosh(x) + q. 
\end{equation*}
In this case, for $q=\theta/(1+\theta)$ the partition function \eqref{9} coincides (up to a numerical factor) with the partition function of the Blume-Capel model.   

\subsection{The result}

For $N\in \mathds{N}$, set $\Theta_N = \{1,2, \dots , N\}$. Then consider
\begin{equation}
 \label{11}
 F_{2N} (x) = \left[\exp\left(\frac{1}{2} \sum_{i,j\in \Theta_{2N}} K_{ij} D_i D_j \right) \prod_{i\in \Theta_{2N}} \phi(x_i) \right]_{{\rm all} \ x_i = x}, 
\end{equation}
where $D_i = \partial /\partial x_i$, $x_i \in \mathds{R}$, cf.  \eqref{5a} and \eqref{9}, and  $\phi$ is as in \eqref{7}. In view of the latter, $F_{2N}$ can be continued to an exponential type entire function of $x\in \mathds{C}$. By Proposition \ref{1pn} it is a Laguerre entire function of $x^2$ whenever all $K_{ij}$ are nonnegative and $\theta \leq 1$.   
\begin{assumption}
 \label{1ass}
 In the statement below, we impose the following conditions on the interaction matrix $K=(K_{ij})$:
 \begin{itemize}
\item[(i)] $K_{ii} = 0$ and  $K_{ij} = K_{ji}\geq 0$ for all $i,j\in \Theta_{2N}$. Moreover, there exists $\varkappa >0$ and a division $\Theta_{2N} = \cup_{k=1}^N \vartheta_k$ into disjoint two-element subsets  $\vartheta_k = \{i_k, j_k\}$ such that $K_{i_k j_k} \geq \varkappa$ for all $k\in \Theta_N$. 
\item[(ii)] There exists a division, $\{\vartheta_k =\{i_k, j_k\}: k\in \Theta_N\}$, of $\Theta_{2N}$ such that, for all $k\in \Theta_N$ and $j\in \Theta_{2N} \setminus \vartheta_k$, the following holds $K_{ji_k} = K_{jj_k}$.
\end{itemize}
\end{assumption}
Condition (i) is not so burdensome. For an appropriate choice of the corresponding finite domain $\Lambda \subset \mathds{Z}^d$,  it is satisfied by standard nearest neighbor interactions on $\mathds{Z}^d$, $d\geq 1$. 
Condition (ii) is more specific -- the just mentioned nearest neighbor interaction fails to meet
it. An example where this condition is satisfied is provided by hierarchical models of Dyson's type, see \cite{KozS} and the references therein. A more detailed discussion of these aspects is given below.
 
Now we are ready to formulate the result.
\begin{theorem}
 \label{1tm}
 Let the matrix $K$ satisfy condition (i) of Assumption \ref{1ass} and $\phi$ be as in \eqref{7} with 
 \begin{equation}
 \label{13}
 \theta \leq\sqrt{\frac{e^\varkappa + e^{-\varkappa} }{2}}.
 \end{equation}
 Then $F_{2N}$ defined in \eqref{11} is an exponential type  entire function possessing imaginary zeros only. 
Furthermore, if $K$ satisfies both conditions of Assumption \ref{1ass} and $\phi$ is such that  
 \begin{equation}
 \label{16a}
 \theta \leq\sqrt{\frac{e^\varkappa + 1 }{2}}.
 \end{equation}
Then $F_{2N}$ has the same property as just mentioned. 
\end{theorem}
Noteworthy, the bound in \eqref{16a} is less restrictive than that in \eqref{13}. This relaxation is achieved by imposing additional restrictions on the elements of $K$.
\noindent
\begin{proof}
As mentioned above,  $F_{2N}$ has the desired property whenever $\theta \leq 1$. Thus, in the proof of both parts of the statement we assume that $\theta >1$. 

By \eqref{7} it readily follows that
\begin{equation}
 \label{15}
 D^n_x \phi(x) = \phi^{(n)} (x)=
 \left\{ \begin{array}{ll}
                  \frac{\cosh x}{1+\theta}, \qquad &{\rm for} \ \ n=2m, \  m\in \mathds{N}, \\[.3cm]
                  \frac{\sinh x}{1+\theta}, \qquad &{\rm for} \ \ n=2m-1, \  m\in \mathds{N}.
                 \end{array} \right. 
\end{equation}
Consider 
\begin{equation}
 \label{14}
\Phi_{\kappa} (x ,y) = \exp\left( \kappa D_x D_y \right) \phi(x) \phi(y), \qquad \kappa>0. 
\end{equation}
By \eqref{15} one gets
\begin{eqnarray}
 \label{16}
 \Phi_{\kappa} (x ,y) & = & \frac{1}{(1+\theta)^2} [ (\theta + \cosh x)(\theta + \cosh y) \\[.2cm] \nonumber & + & \left( \cosh \kappa - 1 \right) \cosh x\cosh y  + \sinh \kappa \sinh x \sinh y ] \\[.2cm] \nonumber & = & \frac{1}{(1+\theta)^2} [ \theta^2 - \cosh \kappa + 2 \theta c_u c_v + e^{\kappa} c_u^2 + e^{-\kappa} c_v^2 ] \\[.2cm] \nonumber & =: & \frac{1}{(1+\theta)^2}\Psi_\kappa (x,y), \quad c_u := \cosh\frac{x+y}{2}, \ \ c_v := \cosh\frac{x-y}{2}.  
\end{eqnarray}
In the proof of both parts of the theorem we write $K = K'' + K'$ with $K'$ chosen specifically for each of the parts. In the proof of the first part, we first pick $\varkappa'\in (0, \varkappa]$ such that 
\begin{equation}
 \label{17}
 \theta^2 = \cosh \kappa,
\end{equation}
which is possible by \eqref{13}. Then we set $K'_{ij} = \kappa$ if $\{ij\} = \vartheta_k$ for some $k\in \Theta_N$, and $K'_{ij}=0$ otherwise. In this case, by condition (i) of Assumption  \ref{1ass} $K''_{ij} \geq 0$ for all $i,j\in \Theta_{2N}$.    
Next we define
\begin{equation}
 \label{18}
G(x_1 , \dots , x_{2N}) = \exp\left( \frac{1}{2} \sum_{i,j \in \Theta_{2N}} K'_{ij} D_i D_j \right) \prod_{i\in \Theta_{2N}}\phi(x_i).
 \end{equation}
Hence, see \eqref{11}  
\begin{equation}
 \label{19}
 F_{2N} (x) = \left[\exp\left(\frac{1}{2} \sum_{i,j\in \Theta_{2N}} K''_{ij} D_i D_j \right) G(x_1 , \dots , x_{2N}) \right]_{{\rm all} \ x_i = x}, 
\end{equation}
and the proof will follow by Proposition \ref{0pn} if we show that $G$ given in \eqref{18} has the corresponding property. 
Denote by $\Phi$ the function as in \eqref{14}, \eqref{16} with $\kappa$ taken as in \eqref{17}. Then the function defined in \eqref{18} can be written as
\begin{equation*}
 G(x_1 , \dots , x_{2N}) = \prod_{k=1}^N \Phi (x_{i_k},x_{j_k}).
\end{equation*}
Thus, $G$ in \eqref{19} has the desired property whenever $\Phi (x,y)$ is nonvanishing for $\Re x >0$, $\Re y >0$. Set $\Psi = (1+\theta)^2 \Phi$. For our choice of $\kappa$, see \eqref{17}, by \eqref{16} we get   
\begin{eqnarray}
 \label{21}
 \Psi (x,y) & = & e^{\kappa} c_u^2+
 e^{-\kappa} c_v^2 + 2 \sqrt{\cosh \kappa}c_u c_v = e^{\kappa}(c_u + \omega_{+} c_v) (c_u + \omega_{+}  c_v),\\[.2cm] \nonumber
 \omega_{\pm} & = & \omega_{\pm} (\kappa) = e^{-\kappa} (\theta \pm \sqrt{\theta^2 - 1})=   e^{-\kappa}  
 \left[\sqrt{\cosh \kappa} \pm \sqrt{2} \sinh \frac{\kappa}{2} \right].
\end{eqnarray}
It is clear that $0< \omega_{-} \leq \omega_{+}$. Let us prove that $\omega_{+}(\kappa) < 1$ for each $\kappa>0$. To this end, we rewrite
\begin{gather}
 \label{21A}
 \omega_{+} (\kappa) = \frac{1}{\sqrt{2}} \left[e^{-\kappa} \sqrt{e^{\kappa}+ e^{-\kappa}}+ e^{-\kappa} \left(e^{\kappa/2} - e^{-\kappa/2}   \right)  \right]. 
 \end{gather}
Since $\omega_{+} (0)=1$, the property in question can be obtained by showing that the derivative of \eqref{21A} satisfies $\omega'(\kappa) < 0$ for $\kappa>0$. Thus,
\begin{eqnarray*}
 \omega'_{+} (\kappa)& =  & -\frac{e^{-\kappa}}{\sqrt{2}}\bigg{[} \sqrt{e^{\kappa}+ e^{-\kappa}} - \frac{1}{2} (e^{\kappa/2} + e^{-\kappa/2}) \bigg{]}\\[.2cm] \nonumber & - & \frac{e^{-\kappa}(e^{\kappa/2}- e^{-\kappa/2})}{\sqrt{2(e^{\kappa}+ e^{-\kappa})}} \bigg{[} \sqrt{e^{\kappa}+ e^{-\kappa}} - \frac{1}{2} (e^{\kappa/2} + e^{-\kappa/2}) \bigg{]}<0,
\end{eqnarray*}
since $\sqrt{e^{\kappa}+ e^{-\kappa}} > \frac{1}{2} (e^{\kappa/2} + e^{-\kappa/2})$ holding for all $\kappa\geq 0$. 
By \eqref{21} one concludes that $\Phi (x,y)$ vanishes if and only if at least one of the following equalities hold:
\begin{eqnarray}
 \label{22}
 c_u + \omega_{+} c_v = 0, \qquad c_u + \omega_{-} c_v = 0, 
\end{eqnarray}
for $0< \omega_{-} <  \omega_{+} < 1$. 
Then
\begin{equation*}
 \varepsilon_{\pm} := \frac{1- \omega_{\pm}}{1+ \omega_{\pm}} \in (0,1),
\end{equation*}
by means of which we rewrite both equations in \eqref{22} in the form
\begin{equation}
 \label{21C}
 \cosh \frac{x}{2} \cosh \frac{y}{2} + \varepsilon_{\pm} \sinh \frac{x}{2} \sinh \frac{y}{2} = 0. 
\end{equation}
For both positive $\Re x$ and $\Re y$, it follows that $|\sinh \frac{x}{2}| >0$, $|\sinh \frac{y}{2}|>0$. Taking this into account we set $ \tanh \frac{y}{2} = \zeta = \rho e^{i\alpha}$. If either of the equalities  in \eqref{22} holds, by \eqref{21C} we get
\begin{equation}
 \label{24}
|e^x|^2 = \frac{1 + \varepsilon^2 \rho^2 -2 \varepsilon \rho \cos \alpha}{1 + \varepsilon^2 \rho^2 +2 \varepsilon \rho \cos \alpha}, \qquad  |e^y|^2 = \frac{1+ \rho^2 + 2 \rho \cos \alpha}{1+ \rho^2 - 2 \rho \cos \alpha },
\end{equation}
with the corresponding choice of $\varepsilon = \varepsilon_{\pm}$.
At the same time, $e^{2 \Re x}= |e^x|^2 >1$ and $e^{2 \Re y}= |e^y|^2 >1$, which contradicts \eqref{24}; hence, neither of the equalities in  \eqref{22} can hold.  This yields the proof of the first part of the theorem.   

Now we choose $K'$ as follows: $K'_{ij} = K_{ij}$ if $\{i,j\} = \vartheta_k$ for some $k\in \Theta_N$, and 
$K'_{ij} = 0$ otherwise. By this choice $K''_{ij} = 0$ if $\{i,j\} = \vartheta_k$ for some $k$. At the same time, by condition (ii) of Assumption \ref{1ass}, for distinct $k, l \in \Theta_N$, it follows that
\begin{equation*}
 \sum_{i \in \vartheta_k} \sum_{j\in \vartheta_l} K''_{ij} D_i D_j = K_{ij} \left(\sum_{i \in \vartheta_k} D_i \right) \left(\sum_{j \in \vartheta_l} D_j \right), 
\end{equation*}
holding for each pair $i\in \vartheta_k$, $j\in \vartheta_l$.
In view of this, we set
\begin{equation}
 \label{27}
\widehat{K}_{kk} =0 \quad {\rm and} \quad    \widehat{K}_{kl} = K''_{ij} = K_{ij} , \quad {\rm for} \ \ i\in \vartheta_k, \ j\in \vartheta_l, \ k\neq l.
\end{equation}
Let $\widehat{\Phi}_k$ be as in \eqref{16} with $\kappa = K_{ij}$ and $\{i,j\} = \vartheta_k$. 
Then the following holds, see \eqref{27},
\begin{eqnarray}
 \label{26}
 \widehat{G} (x_1 , \dots, x_n) & := & \exp\left(\frac{1}{2} \sum_{i,j\in \Theta_{2N}} {K}_{ij} D_i D_j \right) \prod_{i\in \Theta_{2N}} \phi(x_i) \\[.2cm] \nonumber & = & \exp\left(\frac{1}{2} \sum_{k,l\in \Theta_{N}} \widehat{K}_{kl} \left(\sum_{i \in \vartheta_k} D_i \right) \left(\sum_{j \in \vartheta_l} D_j \right) \right) \prod_{k\in \Theta_{N}} \widehat{\Phi}_k(x_{i_k},x_{j_k}) ,
 \end{eqnarray}
with $\vartheta_k=\{i_k, j_k\}$. At the same time, by \eqref{11} we have
\begin{equation}
 \label{28}
 F_{2N} (x) = [\widehat{G} (x_1 , \dots, x_{2N})]_{{\rm all} \ x_i =x} = [\widetilde{G} (y_1 , \dots, y_{N})]_{{\rm all} \ y_k =x},
\end{equation}
where $\widetilde{G}$ is obtained from $\widehat{G}$ by setting $x_{i_k} = x_{j_k} = y_k$ for $\vartheta_k=\{i_k, j_k\}$. By \eqref{26} we then get
\begin{equation}
 \label{29}
 \widetilde{G} (y_1 , \dots, y_{N}) = \exp\left(\frac{1}{2} \sum_{k,l\in \Theta_{N}} \widehat{K}_{kl} D_k D_l \right)\prod_{k\in \Theta_N} \widehat{\Phi}_k (y_{k},y_{k}). 
\end{equation}
By \eqref{16} for $\vartheta_k =\{i,j\}$ it follows that
\begin{eqnarray*}
 \widehat{\Phi}_k (y,y) = \frac{1}{(1+\theta)^2}\widehat{\Psi}_k(y) = \frac{1}{(1+\theta)^2} \left[e^{K_{ij}} \cosh^2 y + 2 \theta \cosh y + \theta^2 - \sinh K_{ij} \right]  .
\end{eqnarray*}
By solving the corresponding quadratic equation one immediately gets that $\widehat{\Psi}_k(y)=0$ if and only if 
\begin{equation}
 \label{31}
 \cosh y = e^{-K_{ij}}(-\theta \pm \delta_{ij}) =:\varepsilon^{\pm}_{ij} ,  
\end{equation}
where
\begin{equation}
 \label{32}
 \delta_{ij} = \sqrt{(e^{K_{ij}} -1) \left[\frac{e^{K_{ij}}+1}{2} - \theta^2 \right]},
\end{equation}
which is real in view of \eqref{16a} and condition (i) of Assumption \ref{1ass}. Let us prove that $|\varepsilon^{\pm}_{ij}|< 1$; recall that $K_{ij} \geq \varkappa >0$, see condition (i). Clearly, it is enough to check this for  $|\varepsilon^{-}_{ij}|$ only. By \eqref{31} and  \eqref{32} one gets
\begin{gather}
 \label{33}
 e^{K_{ij}} (1 - |\varepsilon^{-}_{ij}|)= e^{K_{ij}} (1 +\varepsilon^{-}_{ij})= e^{K_{ij}} - \theta - \delta_{ij}. 
\end{gather}
At the same time
\[
 (e^{K_{ij}} - \theta)^2 - \delta^2_{ij} = \frac{1}{2}\left( e^{K_{ij}} -1\right)^2 + e^{K_{ij}} (\theta+1)^2,
\]
which by \eqref{33} yields $|\varepsilon^{+}_{ij}|< |\varepsilon^{-}_{ij}|< 1$. Thereby, all $y$'s that satisfy \eqref{31} lie on the imaginary axis. By \eqref{29} and \eqref{28} this completes the whole proof.
\end{proof}

\subsection{Corollaries and comments}

For the Blume-Capel model, see \eqref{6}, our Theorem \ref{1tm} yields the following.
\begin{corollary}
 \label{1co}
Let the interaction matrix $J= (J_{ij})$, $i,j \in \Theta_{2N}$ satisfy condition (i) of Assumption \ref{1ass} with certain $\varkappa>0$. Then the partition function $Z_{2N}(h)$ defined in \eqref{3} with $\phi$ as in \eqref{7} has purely imaginary zeros at all $\beta$ (i.e., is as in  \eqref{1}) if $\Delta< \varkappa/2$.     
\end{corollary}
Note that a priori the Lee-Yang property is guaranteed for $\Delta$ satisfying \eqref{8}. 
Instead of this, our Theorem \ref{1tm} yields the following $\beta$-dependent bound for $\Delta$
\[
 \Delta \leq \beta^{-1}\left[\ln 2 + \ln (e^{\beta \varkappa}+ e^{-\beta \varkappa})\right]/2 < \varkappa/2.
\]
In the case of the dilute $s=\pm 1$ Ising model, Theorem \ref{1tm} yields the following. 
\begin{corollary}
 \label{2co}
 Let the interaction matrix $J= (J_{ij})$, $i,j \in \Theta_{2N}$ of the site-thinned (dilute) $s=\pm 1$ Ising model satisfy condition (i) of Assumption \ref{1ass} with certain $\varkappa>0$. Then the annealed partition function of this model, $Z_{2N}(h)$, has purely imaginary zeros if the thinning probability $q$ satisfy
 \[
  \frac{q}{p} = \frac{q}{1-q} \leq \sqrt{(e^{\beta \varkappa}+ e^{-\beta \varkappa})/2}.
 \]
That is, the property in question takes place for each $q$ and sufficiently low temperatures. 
 \end{corollary}
Now let us make additional comments on Assumption \ref{1ass}. The matrix $K$ defines a graph of order $2N$ with vertex set $V=\Theta_{2N}$ and the edges $E = \{\{i,j\}: K_{ij} >0\}$. Condition (i) means that $G=(V,E)$ admits perfect matchings (or `dimer covering', in the terminology of \cite{HL}) such that their elements satisfy $K_{ij}\geq \varkappa>0$.  It is clear that the corresponding models living on $\mathds{Z}^d$, $d\geq 1$ with ferromagnetic nearest neighbor interactions with intensity $J$ satisfy (i) with $\varkappa = J$, cf. Corollary \ref{1co}.  By \eqref{13} $\varkappa$ determines the bound for $\theta$ to which     
it can be continued from $[0,1]$ without affecting the Lee-Yang property of the corresponding model. Note however that interactions of the Curie-Weiss type $J_{ij} = \varkappa/2N$ satisfy (i) with an $N$-dependent $\varkappa$, which makes the statement of Theorem \ref{1tm} trivial in the large $N$ limit.

Condition (ii) is met by hierarchical models of Dyson's type, see, e.g., \cite{BM}. To define such a model one takes $N= 2^n$, and then introduces a family of dimer coverings $\vartheta^{(m)} = \{\vartheta^{(m)}_k: k \in \Theta_{2^{n-m}}\}$, $m=1, \dots n$, where $\vartheta^{(m)}$ is a dimer covering of $\Theta_{ 2^{n-m+1}}$. This family defines the hierarchy of subsets $\Lambda^{(m)}_k \subset\Theta_{2^n}$ in the following way: $\Lambda^{(1)}_k = \vartheta^{(1)}_k$, $k \in \Theta_{2^{n-1}}$, and 
\[
 \Lambda^{(m)}_k = \Lambda^{(m-1)}_r \cup \Lambda^{(m-1)}_s \qquad {\rm for} \ \   \vartheta^{(m)}_k = \{r,s\}, \ \  m\geq 2. 
\]
Then one introduces positive parameters $\varkappa^{(m)}$ and sets $J_{ij} = \varkappa^{(m)}$ if $i$ and $j$ belong to some   
 $\Lambda^{(m)}_k$ but do not belong to one and the same $\Lambda^{(m-1)}_l$ for any $l$. This hierarchical structure allows one to express   
$Z_{\Theta_{ 2^{n-m}}}(h)$ directly through $Z_{\Theta_{2^{n-m+1}}}(h)$, cf. \cite[Eq. (1.11)]{KozS} or \cite[Eq. (3.1)]{KozR}, which essentially facilitates studying the model. In particular, in \cite[Theorem 3.1]{KozR} the following was proved. Given $\nu\in \mathds{N}$, let 
$\phi:\mathds{R}^\nu \to \mathds{R}$ be as in \eqref{5} where $x^2=x\cdot x$ is the scalar product in $\mathds{R}^\nu$. Let also $Z_N$, $N = 2^{n}$ be as in \eqref{3} with
\[
 D_iD_j = \sum_{\iota=1}^\nu \frac{\partial^2}{\partial x_i^\iota \partial x_j^\iota}, 
\]
and positive $J_{ij}$ defined by a hierarchical structure as just described. Then \emph{for all} $\nu$, $Z_N$ satisfies \eqref{1}, where $h^2$ is the corresponding scalar quadrat. In the general case, a similar property was proved only for $\nu=1,2$, see \cite{LiebS}.

\section*{Statemets and Declarations}
\begin{itemize}
 \item 
No funding was received for conducting this study.

 \item 
The author has no competing interests to declare that are relevant to the content of this article.

\end{itemize}

\end{document}